\documentstyle[12pt,aaspp4]{article}

\def\ltsima{$\; \buildrel < \over \sim \;$}
\def\lsim{\lower.5ex\hbox{\ltsima}}
\def\gtsima{$\; \buildrel > \over \sim \;$}
\def\gsim{\lower.5ex\hbox{\gtsima}}
\newcommand{\simgt}{\lower.5ex\hbox{$\; \buildrel > \over \sim \;$}}
\newcommand{\simlt}{\lower.5ex\hbox{$\; \buildrel < \over \sim \;$}}
\newcommand{\bm}[1]{\mbox{\boldmath$#1$}}

\newcommand{\skaco}[1]{\langle{#1}\rangle}
\def\bx#1{\leavevmode\thinspace\hbox{\vrule\vtop{\vbox{\hrule\kern1pt
\hbox{\vphantom{\tt/}\thinspace{\bf#1}\thinspace}}
\kern1pt\hrule}\vrule}\thinspace}

\begin{document}

\title{Implication of $\Omega_m$
through 
the Morphological Analysis \\ of Weak Lensing Fields}
\author{Jun'ichi Sato$^{1}$, Masahiro Takada$^{1}$, Y. P. Jing$^{2}$
and Toshifumi Futamase$^{1}$} 
\affil{{}$^{1}$Astronomical Institute, Tohoku University, Sendai 980-8578, 
Japan\\ {}$^{2}$Shanghai Astronomical Observatories,
the Partner Group of MPI f\"ur Astrophysik,
Nandan Road 80, 200030 Shanghai, China}

\begin{abstract}
We apply the morphological descriptions of two-dimensional 
contour map, the so-called Minkowski functionals (the area fraction, 
circumference, and Euler characteristics), to the 
convergence field $\kappa(\bm{\theta})$ of the large-scale structure 
reconstructed from the shear map produced by the ray-tracing 
simulations. The perturbation theory of 
structure formation has suggested that the non-Gaussian features on 
the Minkowski functionals with respect to the threshold in the 
weakly nonlinear regime
are induced by the three skewness parameters of $\kappa$
that are sensitive to the density parameter of matter, 
$\Omega_{\rm m}$. We show that, in the absence of noise 
due to the intrinsic ellipticities of source galaxies with which 
the perturbation theory results can be recovered,
the accuracy of $\Omega_{\rm m}$ determination is 
improved by $\sim 20\%$ using the Minkowski functionals compared to
the conventional method of using the direct measure of skewness. 

\end{abstract}


\section{Introduction}

Weak gravitational lensing caused by the large-scale structure (LSS) 
of the universe distorts
the images of distant galaxies.
This phenomenon is the so-called {\em cosmic shear}, 
which offers us the unique opportunity to 
measure directly the projected power 
spectrum of dark matter fluctuations 
regardless of the relation between 
dynamical states of the dark matter and luminous matter 
(\cite{Blandford}; \cite{Miralda}; \cite{Kaiser92}).
Recently, several independent measurements of cosmic shear
have been made from deep 'blank-field' CCD imaging surveys,
and reported significant detections of shear variance 
(\cite{Ludvic00}; \cite{Wittman}; \cite{Bacon}; 
\cite{KNL00}; \cite{Maoli}).

Due to the nonlinear evolution of density fluctuation field in the 
large-scale structure, 
the cosmic shear field on small angular scale is 
expected to display significant non-Gaussian features.
Even in this case, for the second moment analysis it has been 
shown that the numerical results from the ray-tracing simulations 
are in remarkably good agreements with the theoretical predictions
using the fitting formula for the nonlinear matter power
spectrum (\cite{JSW}; \cite{Hamana}). On the other hand,
the higher order statistics can provide additional cosmological 
information associated with the non-Gaussian features.
Especially, the normalized skewness 
parameter of the convergence field can be a sensitive 
indicator of the density parameter of matter, $\Omega_{m}$ 
(\cite{BWM}).
However, unfortunately the highly nonlinear evolution 
of third order statistics cannot be simply described by the fitting 
formula for the nonlinear power spectrum alone.
Recently, the extended
method that allows us to perform 
the skewness calculations in the strongly 
nonlinear regime has been developed using ``hyper-extended perturbation 
theory'' (HEPT) (\cite{Hui}; \cite{Scoccimarro99}). 
Nevertheless, several works 
using the ray-tracing simulations have revealed that 
the value predicted by HEPT at relevant angular scales 
does not agree so well with 
the numerical results of skewness parameter 
(\cite{JSW}; \cite{White}; \cite{Hamanab}).
Moreover, we would like to stress that it is difficult 
to have a physical meaning for the fitting formula 
beyond an empirical one. 
Therefore, it will be worth exploring again a new method 
to effectively extracting the non-Gaussian features of the convergence 
field in the weakly nonlinear regime based on the perturbation 
theory, which relies on a more firm physical basis of the structure 
formation. 

A possible method we propose is to use the Minkowski 
functionals with respect to level threshold; this is motivated by
the fact that the functionals give the complete morphological 
descriptions of a considered field (\cite{SB}). 
For a two-dimensional case, the Minkowski functionals consist of 
the area fraction, circumference and Euler characteristics of the 
isocontour curves, where the Euler characteristics is equivalent to
the genus statistics often 
used in the cosmology (\cite{Gott86}). Recently, Matsubara \& Jain
(2000) applied the genus curve to the convergence field reconstructed
from the ray-tracing simulations, and found that the nonlinear 
evolution of convergence induces a deviation from the specific curve of
genus for the Gaussian case. 
On the other hand, the theoretical predictions based on 
the perturbation theory have shown that 
the non-Gaussian features on the Minkowski functionals
are completely characterized by the skewness parameters of the
convergence field in the weakly nonlinear regime 
(\cite{Matsu00} and see also equation (\ref{eqn:mink})). 
These results offer a possibility to extract the skewness parameters
using the Minkowski functionals of the reconstructed convergence field.
The purpose of this Letter 
is thus to investigate how accurately $\Omega_{\rm m}$ can be 
determined from the skewness parameters estimated by fitting 
the numerical results to theoretical predictions of
the Minkowski functionals. 

\section{The Ray-Tracing Simulation and the Minkowski Functionals}
\label{method}
We use shear and convergence fields modeled from the ray tracing 
simulations through the dark matter distribution of N-body 
simulations following the previous methods 
by \cite{Hamana} and \cite{White}.  
The original N-body simulations of the large-scale 
structure were performed with the P$^{3}$M code (see
\cite{Jing94} and \cite{Jing98} in detail).
The following discussions focus on two cosmological models, 
summarized in Table 1,
and we used three different realizations for each model.
As for the power spectrum of matter fluctuations, we assume
the cold dark matter (CDM) model with the transfer function given by 
Bardeen et al. (1986) and the shape parameter $\Gamma=\Omega_{0}h$. 
All the simulations employ $256^{3}$ ($\approx17$ million) particles
in a ($100h^{-1}$Mpc)$^{3}$ comoving box and start at redshift
$z_{i}=36$. The gravitational softening length $\epsilon$ is 
$39h^{-1}$kpc.

We use the multiple lens-plane algorithm to follow the propagations 
of light rays through the simulated matter distributions.
In this algorithm, the matter content of each box 
at a certain redshift is projected onto a single plane perpendicular 
to the line of sight. We use typically
$\sim20$ equally spaced lens-planes in the comoving distance
between source and observer.
The particle positions on each plane are interpolated onto
a grid of size $2048^{2}$.
In order to avoid possible correlations between different lens-planes,
in each plane 
we choose one of the three realizations at the considered redshift and  
then project the mass distribution along a randomly chosen 
one of the three coordinate axes,
translate the mass distribution by a random vector, 
and randomly rotate it in a unit of $\pi/2$.
We consider a set of lens-planes between the source and 
observer as a different realization
and use ten such realizations to estimate the
cosmic variance associated with the measurements of weak lensing fields. 
Further details of the ray-tracing simulation
are given in Hamana, Martel \& Futamase (2000).

\begin{figure*}[t]
\vspace{4cm}
\caption{The reconstructed convergence maps
used to compute the Minkowski functionals
for a field of $3^{\circ}$ on a side.
Two cosmological models are shown:
SCDM model (left) and $\Lambda$CDM model (right).
}
\includegraphics{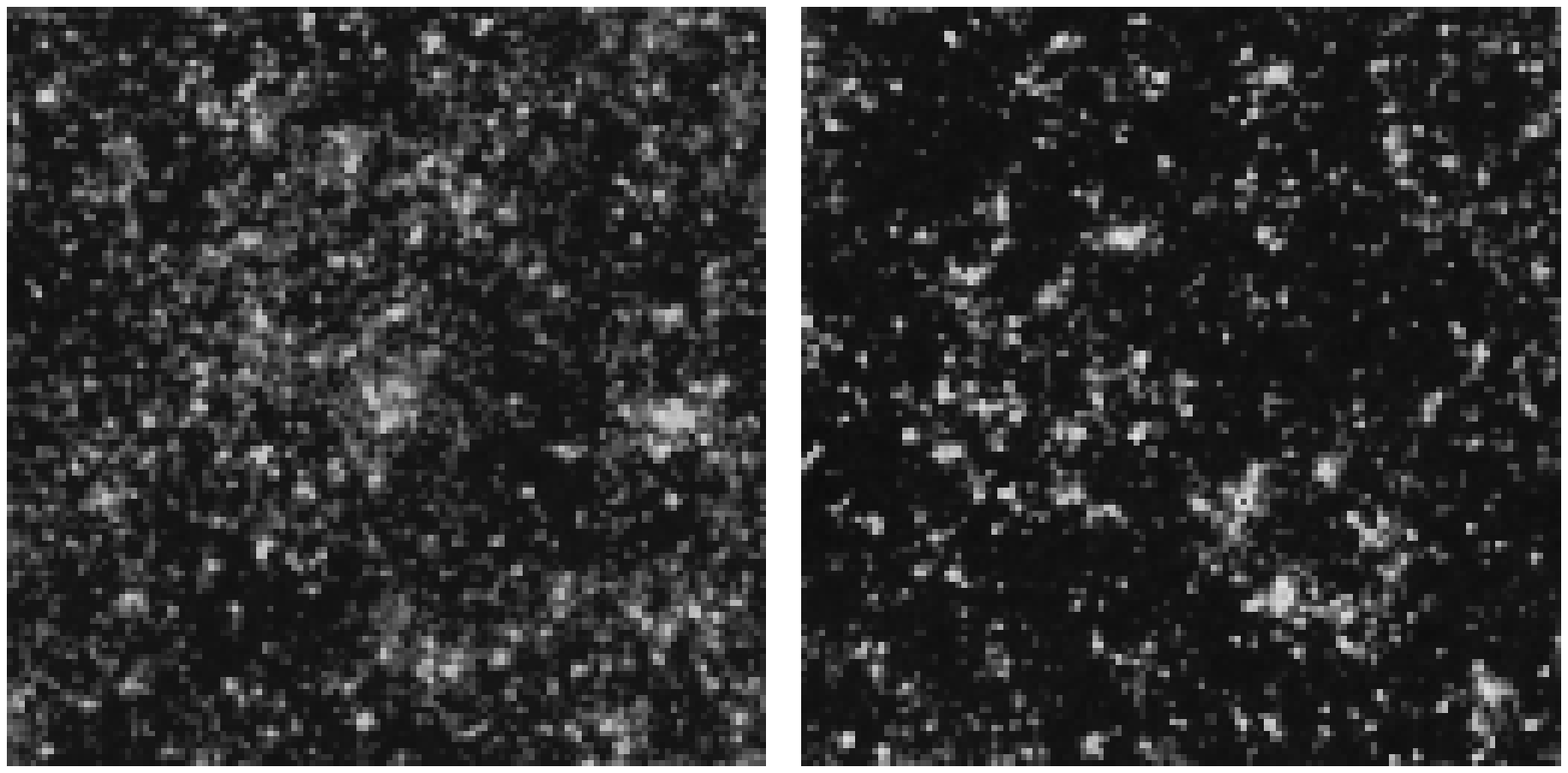}
\label{kappamap}
\end{figure*}

The fields we use are $3^{\circ}$ on a side.
Each light ray is traced by the Born approximation and hence can be
handled as a straight line that radially extends from observer.
Throughout this Letter, we assume that all source galaxies
are at a redshift of $z_{s}=1$ and that their number densities is 
$n=30 \rm{~arcmin}^{-2}$.  We then make the cosmic shear field,
$\gamma(\bm{\theta})$, on each 
grid from the ray-tracing simulations, and perform the smoothing on 
$\gamma(\bm{\theta})$ by using a top-hat filter. 
Using the relation between the Fourier transforms
$\kappa(\bm{\theta})$ and $\gamma(\bm{\theta})$,
$
\hat{\kappa}(\mbox{\boldmath$l$})
= [(l_{1}^{2}-l_{2}^{2})\hat{\gamma}_{1}(\mbox{\boldmath$l$})
+ 2l_{1}l_{2}\hat{\gamma}_{2}(\mbox{\boldmath$l$})]/l^2,
$
and assuming the periodic boundary condition, 
$\kappa(\bm{\theta})$ is reconstructed 
on each grid from the cosmic shear field. 
Figure \ref{kappamap} shows examples of the reconstructed 
convergence field.
To compute the Minkowski functionals,
we label the convergence field by the threshold value $\nu(\bm{\theta})$ 
that is defined by $\nu(\bm{\theta})\equiv
\kappa(\bm{\theta})/\sigma_{0}$, where $\sigma_0$ is the rms 
of $\kappa$ defined by $\sigma^2_0\equiv\skaco{\kappa^2}$.

In a two-dimensional case, the Minkowski functionals are 
the area fraction $v_{0}(\nu)$, circumference $v_{1}(\nu)$, and
Euler characteristics $v_{2}(\nu)$ for the isocontour curve with 
threshold $\nu$ that fully characterize the morphology of 
the field. 
The Euler characteristic is a purely
topological quantity, which  
is equal to the number of isolated high-threshold 
regions minus the number of isolated low-threshold regions. 
To calculate the Minkowski functionals for the
reconstructed convergence field given as pixel data,
we employed the method developed by Winitzki \& Kosowsky (1997).

On the other hand, 
under the hypothesis that the initial perturbations are Gaussian
as supported by the inflationary scenario, 
Matsubara (2000) recently derived the 
analytical formula of the Minkowski functionals based on 
the perturbation theory that can be applied to 
the weakly nonlinear convergence field:
\begin{eqnarray}
v_{0}(\nu)&=&\frac{1}{2}\mbox{erfc}\left( \frac{\nu}{\sqrt{2}} \right)
+ \frac{1}{6 \sqrt{2 \pi}}\mbox{e}^{-\nu^{2}/2}
\sigma_{0}s_{0}\mbox{H}_{2}(\nu),\nonumber \\
v_{1}(\nu)&=&\frac{1}{8\sqrt{2}}\frac{\sigma_{1}}{\sigma_{0}}
\mbox{e}^{-\nu^{2}/2}\left\{1+\sigma_0\left(
\frac{s_{0}\mbox{H}_{3}(\nu)}{6}+\frac{s_{1}\mbox{H}_{1}(\nu)}{3}
\right)\right\},\nonumber\\
v_{2}(\nu)&=&\frac{1}{2(2 \pi)^{\frac{3}{2}}}
\frac{\sigma_{1}^{2}}{\sigma_{0}^{2}}
\mbox{e}^{-\nu^{2}/2}
\left\{ \mbox{H}_{1}(\nu)+\sigma_{0}\left(
\frac{s_{0}\mbox{H}_{4}(\nu)}{6}
+\frac{2s_{1}\mbox{H}_{2}(\nu)}{3}
+\frac{s_{2}}{3} 
\right)\right\},
\label{eqn:mink}
\end{eqnarray}
where $\sigma_{1}$ is defined by
$\sigma^2_1\equiv\skaco{(\nabla\kappa)^2}$ and
H$_{n}(\nu)$ is the $n$th order Hermite polynomial. 
$s_{0}$, $s_{1}$, and $s_{2}$ denote the skewness
parameters defined by $s_0\equiv\skaco{\kappa^3}/\sigma_0^4$, 
$s_1\equiv-(3/4)\skaco{\kappa^2(\nabla^2\kappa)}/(\sigma_0^2\sigma_1^2)$, 
and $s_2\equiv-3\skaco{(\nabla\kappa\cdot\nabla\kappa)(\nabla^2\kappa)}
/\sigma_1^4$, respectively, where the quantity $s_{0}$ is the 
skewness parameter conventionally used in the previous works of 
weak lensing.
Equation (\ref{eqn:mink}) indicates that 
those skewness parameters can be new statistical indicators 
of the deviations from the specific Gaussian predictions of
$v_0(\nu)$, $v_1(\nu)$ and $v_2(\nu)$ with $s_0=s_1=s_2=0$. 
It should be noted that $s_0$, $s_1$ and $s_2$ themselves can be 
given as functions of the cosmological parameters for the 
CDM model and the smoothing scale of the top-hat filter (\cite{BWM}), 
which reveals that the skewness parameters are particularly sensitive 
to $\Omega_m$.
Therefore, we propose that comparing the theoretical predictions 
(\ref{eqn:mink}) to their numerical (or observed) results could place 
constraints on the cosmological parameters. 
In some previous work 
(e.g., \cite{MJ}), the area fraction to labeling the Minkowski 
functionals has been used instead of the threshold
in order to cancel out the horizontal 
shift of those functionals that is due to the nonlinear evolution 
of the underlying density fluctuations on the high threshold side.
However, this operation merely means that the area function $v_0(\nu)$ 
for the non-Gaussian field is transformed closer to its specific 
curve for the Gaussian case. For this reason, we do not employ 
the operation and simply use the threshold $\nu$.

\section{Results}
\label{results}

\begin{figure*}[t]
\vspace{8.8cm}
\caption{Minkowski functionals, the area fraction (left),
circumference (middle),
and Euler characteristics (right)
per square arcmin for the reconstructed convergence fields
with two different smoothing scales of
$\theta =1'$ (upper) and $\theta =8'$ (lower),
respectively. 
Solid (SCDM model) and dashed ($\Lambda$CDM)
curves show the analytical predictions 
of Minkowski functionals calculated by equation (\ref{eqn:mink}).
The mean values (circles and boxes for SCDM and $\Lambda$CDM models,
respectively) and the error bars in each bin of $\nu$ 
are estimated from the ten different realizations with the area of 
$3 \times 3$ square degrees.
\label{fig:mink}
}
\includegraphics{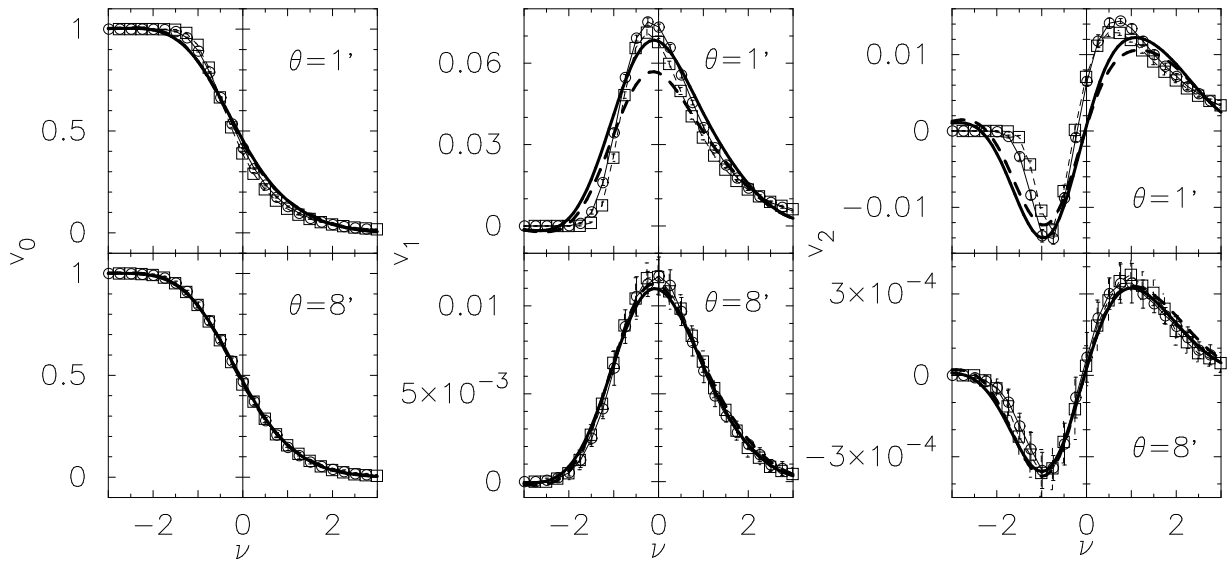}
\label{figv0v1v2}
\end{figure*}

Figure \ref{figv0v1v2} shows both the analytical and numerical results of 
the area fraction $v_{0}(\nu)$ (left panels), circumference $v_{1}(\nu)$ 
(middle panels), and Euler characteristics $v_{2}(\nu)$ (right panels) 
per square arcmin as a function of the threshold $\nu$ for the 
convergence fields with two different smoothing scales of
$\theta=1'$ (upper panels) and $\theta=8'$ (lower panels), respectively.
In those plots, normalizations of the analytical predictions except $v_0(\nu)$
are determined by minimizing the $\chi^{2}$ value for the fitting between 
the predictions (\ref{eqn:mink}) and the numerical results.
The mean values and error bars in each bin of $\nu$ 
are estimated from the ten different realizations with the area of 
$3\times 3$ square degrees, 
and the error corresponds to the cosmic variance 
associated with the measurements of the Minkowski functionals.
Non-Gaussian features 
on the functionals for the noise-free convergence field 
are due to nonlinear gravitational clustering;  at negative
$\nu$ it has a cutoff related
to the minimum $\kappa$ resulting from empty beams
and has a tail at positive $\nu$ due to collapsed halos. 
For the small smoothing scale of $\theta=1'$,
there are large differences between the analytical predictions and the 
numerical results. This is because the highly nonlinear evolution 
of the density field has a large effect on the convergence field. 
For the large smoothing scale of $\theta=8'$, on which the convergence 
field is expected to be in weakly nonlinear regime, 
the numerical results are broadly consistent with the analytical predictions.
Note that the reason that the result of $\theta=8'$ has larger error 
bars than that of $\theta=1'$
is due to the fewer number of statistical samples.

\begin{figure*}[t]
\vspace{8.8cm}
\caption{The directly measured value of skewness parameter (top left)
from the simulated convergence maps,
and the skewness parameters $s_{0}$ (top right),
$s_{1}$ (bottom left), and $s_{2}$ (bottom right)
estimated from fitting between 
the simulated Minkowski functionals and their 
analytical predictions.
In each panel, solid lines with and without circle symbols
represent the results by simulation 
and by the perturbation theory,
respectively,
in SCDM model, while dashed lines with and without box symbols
are results in $\Lambda$CDM model.
The error bars are estimated from ten realizations
for the observation field with $9^{\circ} \times 9^{\circ}$ area
(see text in detail).
In addition, in top left panel
thin lines represent the direct measure of $s_{0}$ in the 
same range of
$-1.5 \le\nu\le 1.5$ as used in the Minkowski functionals method.
\label{fig:skew}
}
\includegraphics{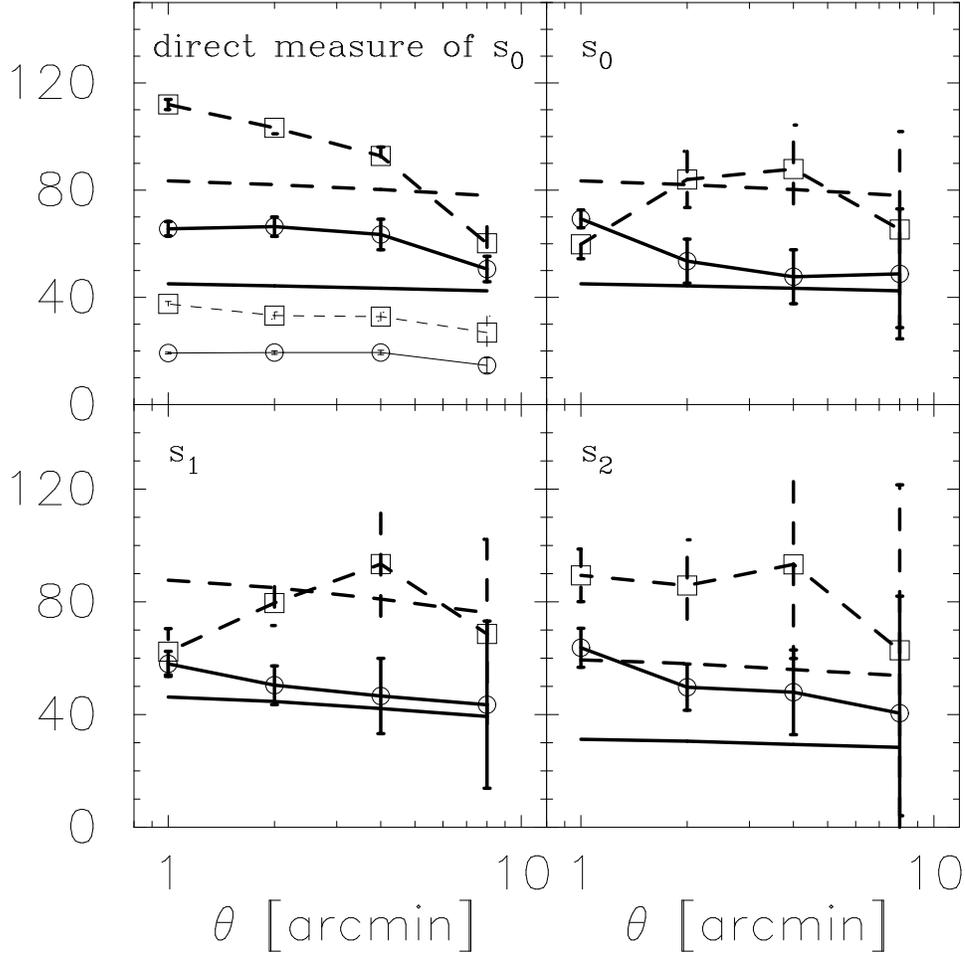}
\label{figparams}
\end{figure*}

Figure \ref{fig:skew} shows the values 
of $s_0$, $s_1$ and $s_2$ calculated by the perturbation theory, 
the direct measurement of $s_0$ (top left) from the reconstructed 
convergence field and the estimations of $s_0$ (top right), $s_1$ (bottom left)
and $s_2$ (bottom right) obtained from 
the $\chi^2$ fitting between the theoretical predictions
(\ref{eqn:mink}) of the Minkowski functionals 
and their simulation results.  
Here we have used only the simulation data in the range of 
$-1.5\le\nu\le 1.5$, because we expect that 
the convergence field in this range is still in the weakly nonlinear 
regime and therefore can be applied to 
the perturbation theory predictions.
In these figures, assuming that the survey of weak lensing is 
performed over the area of $9\times 9$ square degrees, 
we estimated the error bars by multiplying the variance directly 
obtained from 
the ten realizations as shown in Figure \ref{fig:mink} 
by a factor of $1/3$.
We have confirmed that the measurement of Euler characteristics, 
$v_2$, is also sensitive to the discreteness effect of pixel data. 
Therefore, to minimize the unresolved uncertainties, 
we determined the parameters of $s_0$, 
$s_1$, and $s_2$ in the following procedure. 
First, we determine $s_0$ from the fitting of 
$v_0(\nu)$ because the non-Gaussian features of $v_0(\nu)$
in the theoretical prediction (\ref{eqn:mink}) depends on $s_0$ and
$\sigma_0$, where $\sigma_0$ is also computed directly from the
reconstructed convergence field according to the definition
$\sigma_0^2=\skaco{\kappa^2}$.
Similarly, by using the
already determined value of $s_0$,
we determine $s_1$ from the shape of $v_1$. 
Finally, we use the shape of 
Euler characteristics $v_2(\nu)$ to determine the $s_2$ parameter.
Note that this fitting procedure causes the large error of $s_2$.  
The top left panel in Figure \ref{fig:skew} shows that for all 
smoothing scales the direct measurement of $s_0$ tends to largely
overestimate the value of $s_0$ calculated by the perturbation theory.
This is because the direct measurement is more sensitive to the strong 
nonlinear rare events in the convergence distribution such as 
halos of dark matter. 
On the other hand, for SCDM model with $\theta=2',4'$ and $8'$, 
the values of $s_{0}$ obtained from our method using $v_0(\nu)$ 
fairly improve the
estimations for $s_0$ predicted by the perturbation theory. 
For comparison, thin lines in the top left panel of Figure \ref{fig:skew}
show the direct measurement of $s_{0}$
in the same range of $\nu$ ($-1.5 \le\nu\le 1.5$) as used in our
method. It is still clear that the modified 
direct measurement of $s_0$ 
also fails to predict its value from the perturbation theory
for all the smoothing scales.
Similarly, the values of $s_1$ obtained from our method 
are very similar to the values of $s_1$ from the perturbation 
theory for the smoothing scales of $\theta\simgt 2'$.  
However, one can see that the result of $s_2$ from our method
cannot reproduce the value of the perturbation theory mainly
because of the fitting procedure described above, and the results 
of $s_0$ and $s_1$ for the smallest smoothing scale of 
$\theta=1'$ do not work well.
For these reasons, we will not use the results of $s_0$ and 
$s_1$ for $\theta=1'$ and $s_2$  for the determination of $\Omega_m$.
On the other hand, it apparently 
seems that the errors of the skewness determinations for $\Lambda$CDM 
model are larger than those of SCDM. 
This result comes from the fact that 
the skewness variation $\Delta s_0=10$ for the flat universe models 
around $\Lambda$CDM model corresponds to $\mid \Delta \Omega_m \mid=0.05$,  
while around SCDM model $\Delta s_{0}=0.9$
corresponds to the same $\mid \Delta \Omega_m \mid$.  
Actually, as will be shown, the relative accuracy of the 
$\Omega_m$ determination 
is not so different in both SCDM and $\Lambda$CDM models.

Table \ref{tab:om} summarizes the results for the $\Omega_m$
determination with a best-fit value and $1\sigma$ error, 
which are obtained from the direct measurements of $s_0$ 
and from the estimations of $s_0$ and $s_1$ using the Minkowski 
functionals for the smoothing scales of $\theta=2', 4'$ and $8'$,
respectively. 
We here employed the current favored flat universe models 
with $\Omega_m+\Omega_\lambda=1$. 
The table clearly shows that 
our method improves the accuracy of $\Omega_m$ determination
by $\sim 20\%$ compared to
that determined from the direct measure of skewness.

\section{Discussion}

In this Letter we addressed the issue of how
accurately the density parameter, $\Omega_{m}$, can be 
determined from the non-Gaussian signatures 
in the simulated weak lensing field based on the perturbation theory of 
structure formation instead of the empirical fitting formula.
For this purpose,  we have shown that
the Minkowski functionals of convergence maps 
reconstructed from the cosmic shear field can be 
a useful new method.
This is because 
the Minkowski functionals can effectively 
pick up the weakly nonlinear non-Gaussian features in the 
appropriate range of threshold, in which 
the perturbation theory can be safely applied.
In fact, our numerical results have shown that 
the $\Omega_m$ determination of using 
the Minkowski functionals produces $\sim 20\%$ accurate
best-fit value to the input value of $\Omega_m$ compared with the  
result of using the direct measurement of skewness. 
However, we still have to further investigate 
possible uncertainties due to the limited number of numerical 
realizations used in this Letter by increasing the number, and 
this will be our future work.

In this Letter, we have not considered the effect of 
intrinsic ellipticities of source galaxies on our method. 
Nevertheless, for the practical purpose,
it is critical to take into account this effect,
and therefore we will need the theoretical 
predictions of the Minkowski functionals, including the noise effect.
This study is now in progress
and will be presented elsewhere. In practice, it will also be 
necessary to take into account the redshift distribution of 
source galaxies. However, previous works have quantitatively 
shown that, even if using a more realistic model for 
the redshift distribution of source galaxies as expressed by 
$n(z)\propto z^{2}\exp[-(z/z_{0})^{2.5}]$ with the mean redshift of 
unity, the magnitude of cosmic shear signal is changed only by 
$\sim 10\%$ compared to the result of using all the sources distributed 
at $z_s=1$ (e.g., \cite{JSW}). 
Therefore, we prospect that the change of source distribution 
does not largely affect our results.

\section*{Acknowledgments}
The authors would like to thank for T. Hamana, 
K. Umetsu and J. Schmalzing for 
useful discussions and valuable comments. 
M.T. also acknowledges a support from a JSPS fellowship.
Y.P.J. is supported in part by
the One-Hundred-Talent Program and
by NKBRSF (G19990754).

\pagebreak

\begin{table*}
\caption[]{Summary of the parameters used for the N-body
simulations. $h$ is the Hubble constant in units of $100\,{\rm
km\,s^{-1}\,Mpc^{-1}}$. }
\centering
\begin{tabular}{l|*{7}{c}}
\hline
Model & $\Omega_m$ & $\Omega_\Lambda$ & $\Gamma$ & $\sigma_8$ &
m$_{\rm{p}}(h^{-1}M_{\odot})$) \\
\hline
SCDM         & 1.0 & 0.0 & 0.5 & 0.6 & 1.7$\times$10$^{10}$ \\
$\Lambda$CDM & 0.3 & 0.7 & 0.21 & 1.0 & 5.0$\times$10$^{9}$ \\
\hline
\end{tabular}
\label{tab:1}
\end{table*}

\begin{table*}
\caption[]{The value of $\Omega_{m}$ 
estimated from the direct measurement of $s_{0}$
and from the measurements of $s_{0}$ and $s_{1}$ through
the fitting of Minkowski functionals
for simulation data with $\theta =2', 4'$ and $8'$.
We here employed the flat universe models with
$\Omega_{m}+\Omega_{\Lambda}=1$.}
\centering
\begin{tabular}{l|*{2}{c}}
\hline
Model & $\Omega_m$ from the direct measure of $s_0$ & $\Omega_m$ from
Minkowski functionals \\
\hline
SCDM ($\Omega_{m}=1.0$)         & 0.50 $\pm$ 0.19 & 0.78 $\pm$ 0.22 \\
$\Lambda$CDM ($\Omega_{m}=0.3$) & 0.24 $\pm$ 0.05 & 0.31 $\pm$ 0.07 \\
\hline
\end{tabular}
\label{tab:om}
\end{table*}

\end{document}